\documentclass[prl,reprint,a4paper,floatfix,superscriptaddress]{revtex4-1}

\usepackage{graphicx}
\usepackage{amsmath}
\usepackage{hyperref}
\usepackage{pdfpages}

\newcommand{\bog}{B$_{1\textnormal{g}}$ }
\newcommand{\etal}{\textit{et al. }}

\begin{document}

\title{Dielectric versus magnetic pairing mechanisms in high-temperature cuprate superconductors investigated using Raman scattering}
 
\author{B.P.P. Mallett}
\email[Corresponding author, ]{benjamin.mallett@unifr.ch}
\affiliation{MacDiarmid Institute, SPCS, Victoria University P.O. Box 600, Wellington 6140, New Zealand}

\author{T. Wolf}
\affiliation{Karlsruhe Institute of Technology, Postfach 3640, Karlsruhe 76021, Germany}

\author{E. Gilioli}
\affiliation{IMEM-CNR, Institute of Materials for Electronics and Magnetism, 43124 Parma, Italy}

\author{F. Licci}
\affiliation{IMEM-CNR, Institute of Materials for Electronics and Magnetism, 43124 Parma, Italy}

\author{G.V.M. Williams}
\affiliation{MacDiarmid Institute, SPCS, Victoria University P.O. Box 600, Wellington 6140, New Zealand}

\author{A.B. Kaiser}
\affiliation{MacDiarmid Institute, SPCS, Victoria University P.O. Box 600, Wellington 6140, New Zealand}

\author{N.W. Ashcroft}
\affiliation{Laboratory of Atomic and Solid State Physics, Cornell University, Ithaca, New York 14853-2501, USA}

\author{N. Suresh}
\affiliation{MacDiarmid Institute, Industrial Research Limited, P.O. Box 31310, Lower Hutt, New Zealand.}

\author{J.L. Tallon}
\affiliation{MacDiarmid Institute, Industrial Research Limited, P.O. Box 31310, Lower Hutt, New Zealand.}

\date{\today}


\begin{abstract}
We suggest, and demonstrate, a systematic approach to the study of cuprate superconductors, namely, progressive change of ion size in order to systematically alter the interaction strength and other key parameters. $R$(Ba,Sr)$_2$Cu$_3$O$_y$ ($R$=\{La, \ldots Lu,Y\}) is such a system where potentially obscuring structural changes are minimal. We thereby systematically alter both dielectric and magnetic properties. Dielectric fluctuation is characterized by ionic polarizability while magnetic fluctuation is characterized by exchange interactions measurable by Raman scattering. The range of transition temperatures is 70 to 107 K and we find that these correlate only with the dielectric properties, a behavior which persists with external pressure. The ultimate significance may remain to be proven but it highlights the role of dielectric screening in the cuprates and adds support to a previously proposed novel pairing mechanism involving exchange of quantized waves of electronic polarization.

%The standard separation in metals between valence and core states leads in cuprate superconductors to constituent ions which may sustain both charge and spin fluctuations. We change ion size in $R$A$_2$Cu$_3$O$_y$ (A=(Ba,Sr) and $R$=(La, \ldots Lu,Y)) in order to systematically alter both dielectric and magnetic properties. Dielectric fluctuation is characterized by ionic polarizability while magnetic fluctuation is characterized by exchange interactions measurable by Raman scattering. The range of transition temperatures is 70 to 107 K and we find that these correlate only with the dielectric properties, a behavior which persists with external pressure. This highlights the fundamental importance of charge fluctuation and dielectric screening in the cuprates and may signal a novel pairing mechanism having its origin with screened and quantized waves of electronic polarization.
\end{abstract}

\maketitle

\begin{figure}
\centerline{\includegraphics*[width=75mm]{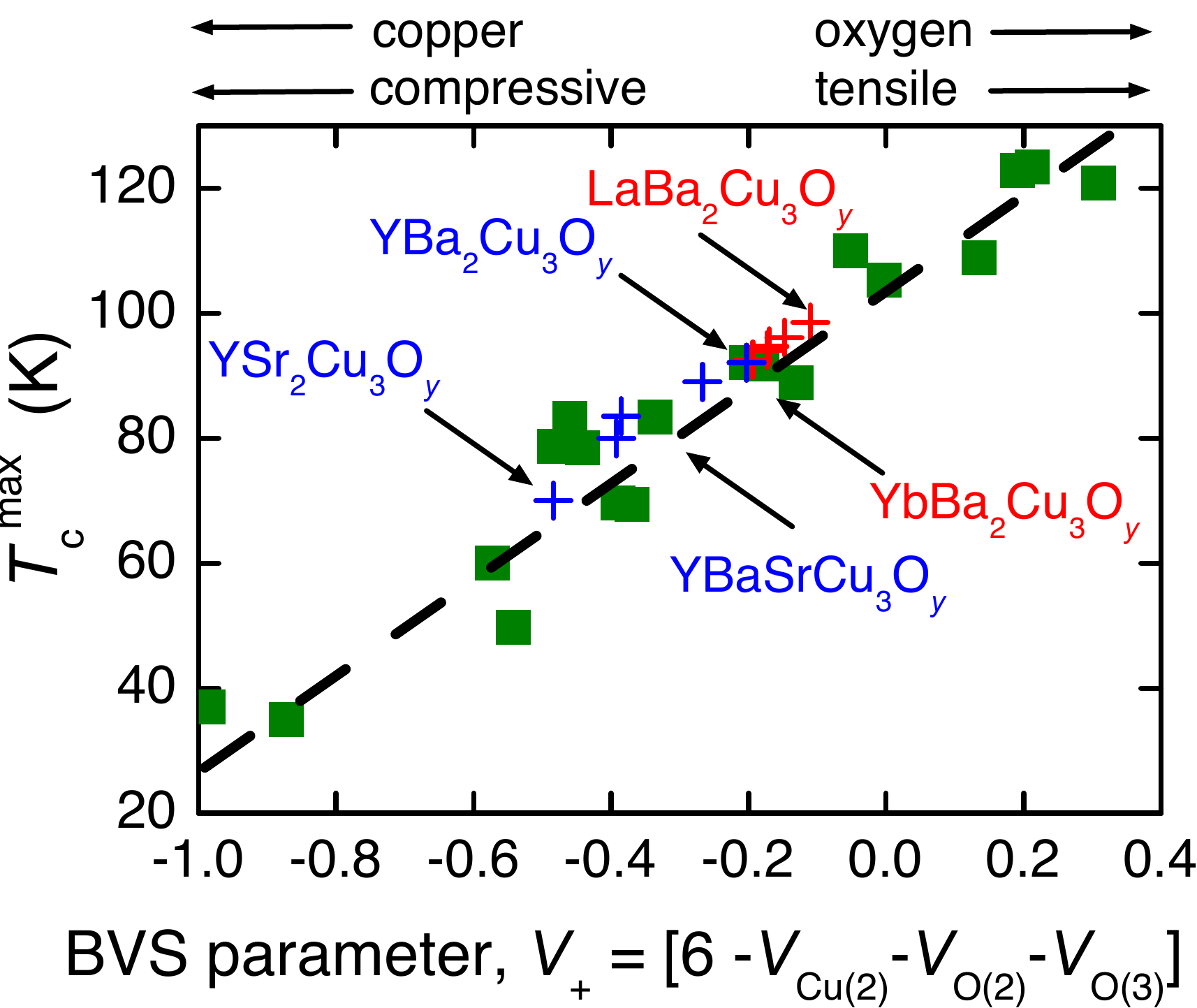}}
\caption{\label{fig:bvs}$T_c^{\textnormal{max}}$, plotted as a function of the bond valence sum parameter $V_+ = 6 - V_{\textnormal{Cu}(2)} - V_{\textnormal{O}(2)} - V_{\textnormal{O}(3)}$. Green squares: as previously reported \cite{tallon1990}; red crosses: $R$Ba$_2$Cu$_3$O$_y$ ($R$ = La, Nd, Sm, Gd, Dy and Yb); blue crosses: YBa$_{2-x}$Sr$_x$Cu$_3$O$_y$ ($x = 0$, 0.5, 1.0, 1.25 and 2).}
\end{figure}

The physical mechanism for electron pairing in cuprate superconductors remains uncertain. While there may be a broad consensus that it is probably magnetic in origin \cite{scalapino2012} a continuing challenge was the apparent low spectral weight of associated spin fluctuations, as measured by inelastic neutron scattering \cite{maksimov2010}. Recent studies using resonant inelastic x-ray scattering (RIXS) seem to locate the missing weight by identifying intense paramagnon excitations across the entire superconducting phase diagram  \cite{letacon2011}. However, there are major material-dependent variations in superconducting properties, as summarized in Fig.~\ref{fig:bvs} (see also \cite{ohta1991}), that remain unexplained and any successful theory must account for these.  The basic Hubbard model can account for the observed generic phase behavior as a function of interaction strength and doping \cite{gull2013} but not for these material-dependent systematics. The problem is exacerbated by the complex and variable structure of the cuprates where some of the behavior may be systematic while some may be attributable to disorder \cite{eisaki2004}, or uncontrolled structural variation in terms of e.g. buckling angles and apical oxygen bond lengths. 

We propose an approach to resolving this impasse by developing a suite of experiments that explore the effects of external pressure and changing ion size (internal pressure) on all the key energy scales using a model system in which structural variables remain essentially unchanged.  In this way the true underlying material-dependent variation might be exposed and at the same time be used to test competing theoretical models. At the very least this offers a way to systematically vary the interaction strength, a probe which has hitherto been missing in experimental studies. We illustrate the approach in the model system $R$(Ba,Sr)$_2$Cu$_3$O$_y$ by studying the systematic variation in the nearest-neighbor exchange interaction $J$, as measurable by Raman two-magnon scattering under changing external and internal pressure. In $R$(Ba,Sr)$_2$Cu$_3$O$_y$ the change in buckling angle is less than $2^{\circ}$ for a given doping \cite{guillaume1994, licci1998} and disorder is essentially absent in the end members. Even for YBaSrCu$_3$O$_y$ we find no suppression of superfluid density as measured by muon spin relaxation \cite{mallettthesis} which might arise from disorder. Moreover, the ion-size-dependent strain tensor for this system reduces to a nearly pure dilatation with almost no tetragonal or orthorhombic distortion; see Supplemental Material \cite{mallettsm}. %Details of standard sample preparation, processing and characterization are given in the Supporting Material \cite{mallettsm}.  

%We find tight constraints are imposed on any model by the fact that internal and external....
We find that internal and external pressures have identical effects on $J$ but opposite effects on $T_c^{\textnormal{max}}$.  Though $T_c^{\textnormal{max}}$ fails to correlate with $J$ it does correlate with the dielectric properties as described by the refractivity sum $\sum_i{n_i \alpha_i}$, where $n_i$ are the ion densities and $\alpha_i$ their polarizabilities, thus highlighting the key role of dielectric screening.  While this work leaves some open questions, the correlations elucidated do warrant further study.  At the least, it does illustrate a necessary and overdue systematic approach to the study of cuprate physics which hitherto has largely been confined to comparative studies of YBa$_2$Cu$_3$O$_y$, Bi$_2$Sr$_2$CaCu$_2$O$_{8+\delta}$ and (La,Sr)$_2$CuO$_4$ - widely differing systems.  The approach can easily be extended to study systematic changes in electronic structure and low-energy excitations in the cuprates. 

%Here we find that $T_c^{\textnormal{max}}$ does not scale with $J$. Instead we find a strong correlation of $T_c^{\textnormal{max}}$ with the dielectric properties as described by the density-weighted sum of the ion polarizabilities, $\alpha_i$ - the so-called refractivity sum, $\sum_i{n_i\alpha_i}$.  Our results could be explicable in a magnetic pairing scenario in terms of effective screening of long-range repulsive interactions \cite{raghu2012optimaltc}. But alternatively, they may signal a previously-proposed novel pairing mechanism where the exchange boson is a coherent quantized wave of electronic polarization \cite{atwal2004}.  Whatever the final outcome, we believe this illustrates 

%We too examine such relationships in a simpler model system, $R$A$_2$Cu$_3$O$_y$ and we measure $J$ directly using two-magnon Raman scattering. But we find quite the opposite result; $T_c^{\textnormal{max}}$ does not scale with $J$. Instead we find a strong correlation with the dielectric properties as described by the density-weighted sum of the ion polarizabilities (the refractivity sum).  Something of this character might yet be explicable in a magnetic pairing scenario in terms of effective screening of long-range repulsive magnetic interactions \cite{raghu2012optimaltc}. But, it may also signal a novel pairing mechanism where the exchange boson is a quantized wave of electronic polarization \cite{atwal2004}.

We are motivated by a central paradox of cuprate physics: external pressure increases $T_c^{\textnormal{max}}$ \cite{schillingchapter}, whereas internal pressure, as induced by isovalent ion substitution, decreases $T_c^{\textnormal{max}}$ \cite{tallon1990, marezio2000}. Figure~\ref{fig:bvs} shows $T_c^{\textnormal{max}}$ plotted against the composite bond valence sum parameter, $V_+ = 6 - V_{\textnormal{Cu}(2)} -V_{\textnormal{O}(2)} -V_{\textnormal{O}(3)}$, taken from Ref.~\cite{tallon1990} (green squares). Here $V_{\textnormal{Cu}(2)}$, $V_{\textnormal{O}(2)}$, and $V_{\textnormal{O}(3)}$ are the planar copper and oxygen BVS parameters and the plot reveals a remarkable correlation of $T_c^{\textnormal{max}}$ across single-, two-, and three-layer cuprates. %For the latter $V_+$ was calculated for the inner of the three layers. 
We may write $V_+ =(2-V_{\textnormal{O}(2)})+(2-V_{\textnormal{O}(3)}) - (V_{\textnormal{Cu}(2)}-2)$ and hence $V_+$ is a measure of doped charge distribution between the Cu and O orbitals \cite{brown1985} but is also a measure of in-plane stress \cite{brown1989}, as noted at the top of the figure. Evidently stretching the CuO$_2$ plane increases $T_c^{\textnormal{max}}$.  However, $V_+$ is a compound measure and also reflects physical displacement of the apical oxygen away from the Cu atoms.

Crucially, this plot reveals that all cuprates follow a systematic behavior. \emph{There are no anomalous outliers}. It is common to regard La$_{2-x}$Sr$_x$CuO$_4$ as anomalous because of its propensity for disorder. But the leftmost data point in Fig.~\ref{fig:bvs} shows that it is entirely consistent with the other cuprates. It remains then to determine just what this $V_+$ parameter encapsulates so systematically.

To this plot we add new data for the superconductors $R$Ba$_{2-x}$Sr$_{x}$Cu$_3$O$_{y}$, as $R$ is varied with $x=0$ (red crosses) and, in the case of $R$ = Y, $x =$ 0, 0.5, 1, 1.25 and 2 (blue crosses). We use the structural refinements of Guillaume \etal \cite{guillaume1994}, Licci \etal \cite{licci1998}  and Gilioli \etal \cite{gilioli2000} and calculate $V_+$ in the same way as previously \cite{tallon1990}. Notably, the global correlation is also preserved across this model system, reflecting the progressive compression of the lattice as ion size decreases and the effective internal pressure increases. Fig.~\ref{fig:bvs} thus summarizes a general feature of the cuprates, namely that internal pressure decreases $T_c^{\textnormal{max}}$ while external pressure increases $T_c^{\textnormal{max}}$ \cite{schillingchapter}. What then is the salient difference between internal and external pressure on $T_c^{\textnormal{max}}$?

The magnitude of $T_c^{\textnormal{max}}$ will be set in part by the energy scale, $\hbar \omega_{\textnormal{B}}$, of the pairing boson and also by $N(E)$, the electronic (DOS).   In the underdoped regime the DOS is progressively depleted by the opening of the pseudogap, whereas the overdoped DOS is enhanced by the proximity of the van Hove singularity (vHs). Here we focus on $\hbar\omega _{\textnormal{B}}$ which in many magnetic pairing models is governed to leading order by $J$ \cite{scalapino2012, letacon2011, kee2002}; see Ref.~\cite{mallettsm} for details. In the one-band Hubbard model, $J=-4t^2/U$, where $t$ is the nearest-neighbor hopping integral and $U$ is the on-site Coulomb repulsion energy. Our question then is, does $T_c^{\textnormal{max}}$ correlate with $J$?

\begin{figure}
\centerline{\includegraphics*[width=75mm]{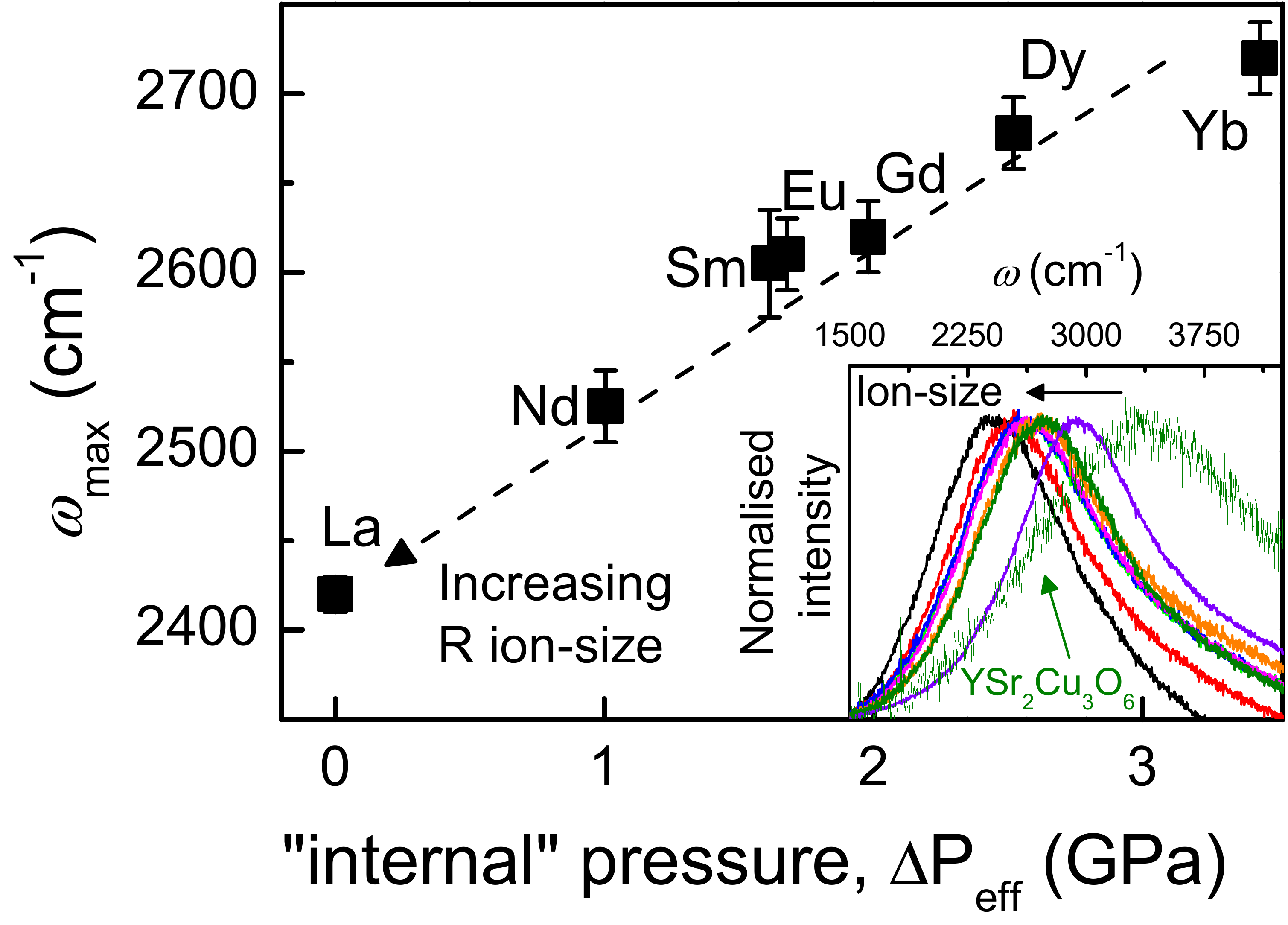}}
\caption{\label{fig:twomag}
The \bog Raman two-magnon peak frequency $\omega_{\textnormal{max}}$ for each $R$Ba$_2$Cu$_3$O$_6$ ($R$-123) sample plotted against the effective internal pressure using $\Delta P_{\textnormal{eff}} = - B.\Delta V/V_0$ where $B= 78.1$ GPa is the bulk modulus \cite{suenaga1991} and $\Delta V = V-V_0$ is referenced to La-123. The inset shows normalized Raman spectra for $R$-123 single crystals and for polycrystalline YSr$_2$Cu$_3$O$_6$.}

\end{figure}

We measured $J$ in single crystals of deoxygenated $R$Ba$_2$Cu$_3$O$_6$ ($R$-123) using \bog Raman scattering where the two-magnon peak occurs at frequency $\omega_{\textnormal{max}}=3.2J/\hbar$ \cite{chubukov1995}.  We choose deoxygenated $R$-123 because the undoped state is the only truly reproducible doping level across the $R$-series and, moreover, the two-magnon peak is not easy to resolve at optimal doping.  Though $J$ is doping dependent we fully expect that it will retain the same systematic variation with ion size reported below for all doping levels. The normalized raw data are shown in the inset of Fig.~\ref{fig:twomag}. As the $R$ ion size decreases, the \bog two-magnon peak shifts to higher energy as expected due to the increased overlap between Cu 3$d$ and O 2$p$ orbitals.

\begin{figure}
\centerline{\includegraphics*[width=75mm]{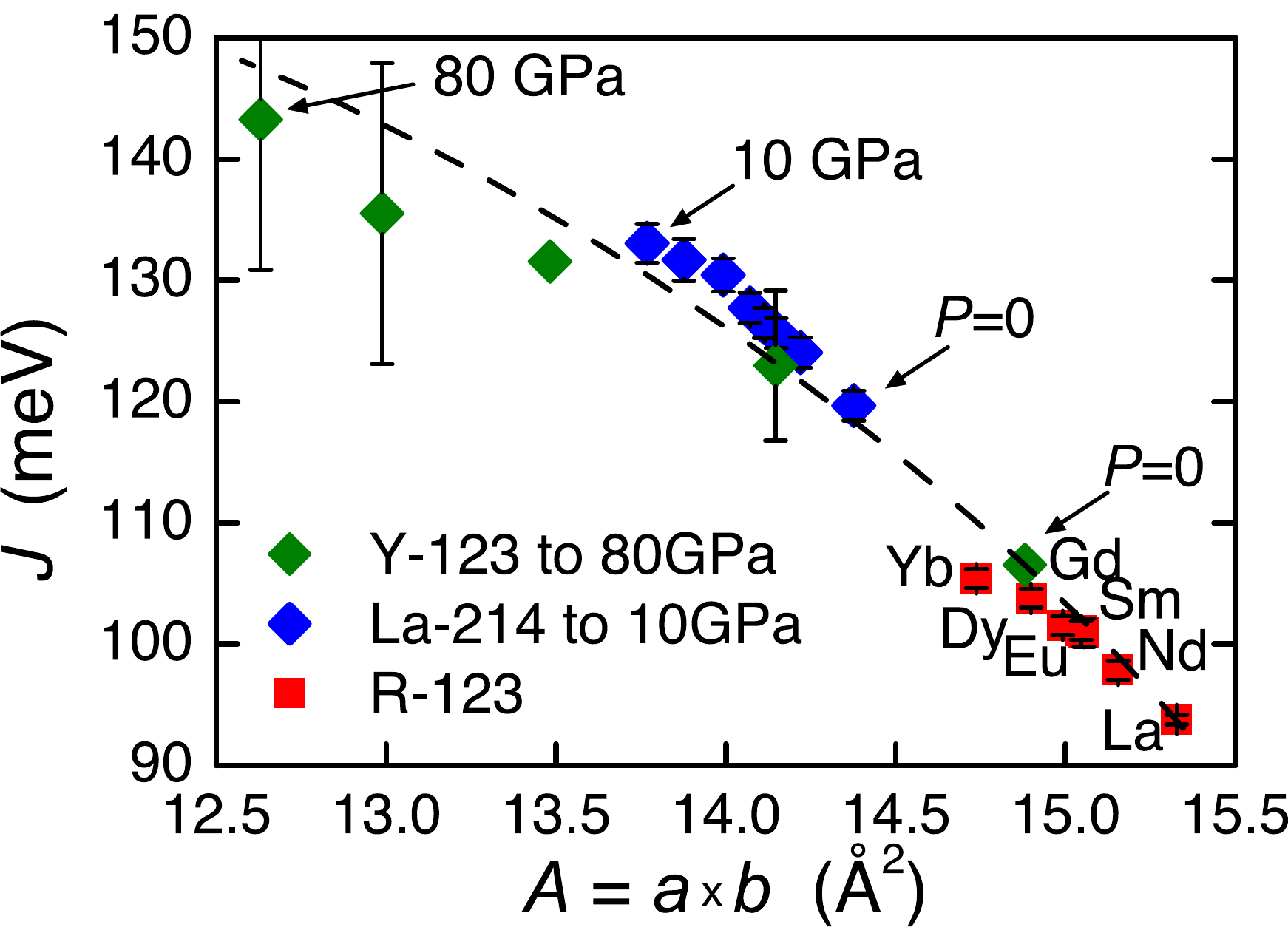}}
\caption{\label{fig:jvsa}The unit-cell basal plane area ($A$) dependence of $J$ determined from two-magnon scattering on $R$Ba$_2$Cu$_3$O$_6$ (red squares) where ``internal pressure'' is the implicit variable. The effect of external pressure \cite{aronson1991} is also shown for La$_2$CuO$_4$ (0 to 10 GPa, blue diamonds) and YBa$_2$Cu$_3$O$_{6.2}$ to 80 GPa \cite{maksimov1994} (green diamonds).} % A single behavior for $J(A)$ is preserved for internal and external pressure.} % Inset: $J$ versus $A$ for pressure-dependent two-magnon data for  compared with our ion-size-dependent data for $R$Ba$_2$Cu$_3$O$_6$ (red squares).}
\end{figure}

The relative shift in effective internal pressure $\Delta P_{\textnormal{eff}}$ may be estimated from the change in volume $\Delta V$ using $\Delta P_{\textnormal{eff}} = -B.\Delta V/V_0$, where $B = 78.1$~GPa is the bulk modulus for deoxygenated YBa$_2$Cu$_3$O$_6$ \cite{suenaga1991} and $\Delta V=V-V_0$ is referenced to La-123.  We plot $\omega_{\textnormal{max}}$ vs $\Delta P_{\textnormal{eff}}$ in Fig.~\ref{fig:twomag}.  Further, the dependence of $J$ on basal area $A$ is plotted in Fig.~\ref{fig:jvsa}. To this we also show in Fig.~\ref{fig:jvsa} the effect of external pressure on $J$ in La$_2$CuO$_4$ \cite{aronson1991} (blue diamonds) ranging up to 10~GPa, as annotated, and for YBa$_2$Cu$_3$O$_{6.2}$ under external pressure up to 80~GPa \cite{maksimov1994} (green diamonds). Bearing in mind the nonlinearity that must occur on approaching 80~GPa, the dependence of $J$ on $A$ is similar across the entire range, irrespective of whether the pressure is {\it internal} or {\it external} in origin.  We expect this uniform dependence to be preserved at optimal doping. This is our first main result and it contrasts the opposing effects of internal and external pressure on $T_c^{\textnormal{max}}$. 

\begin{figure}
\centerline{\includegraphics*[width=75mm]{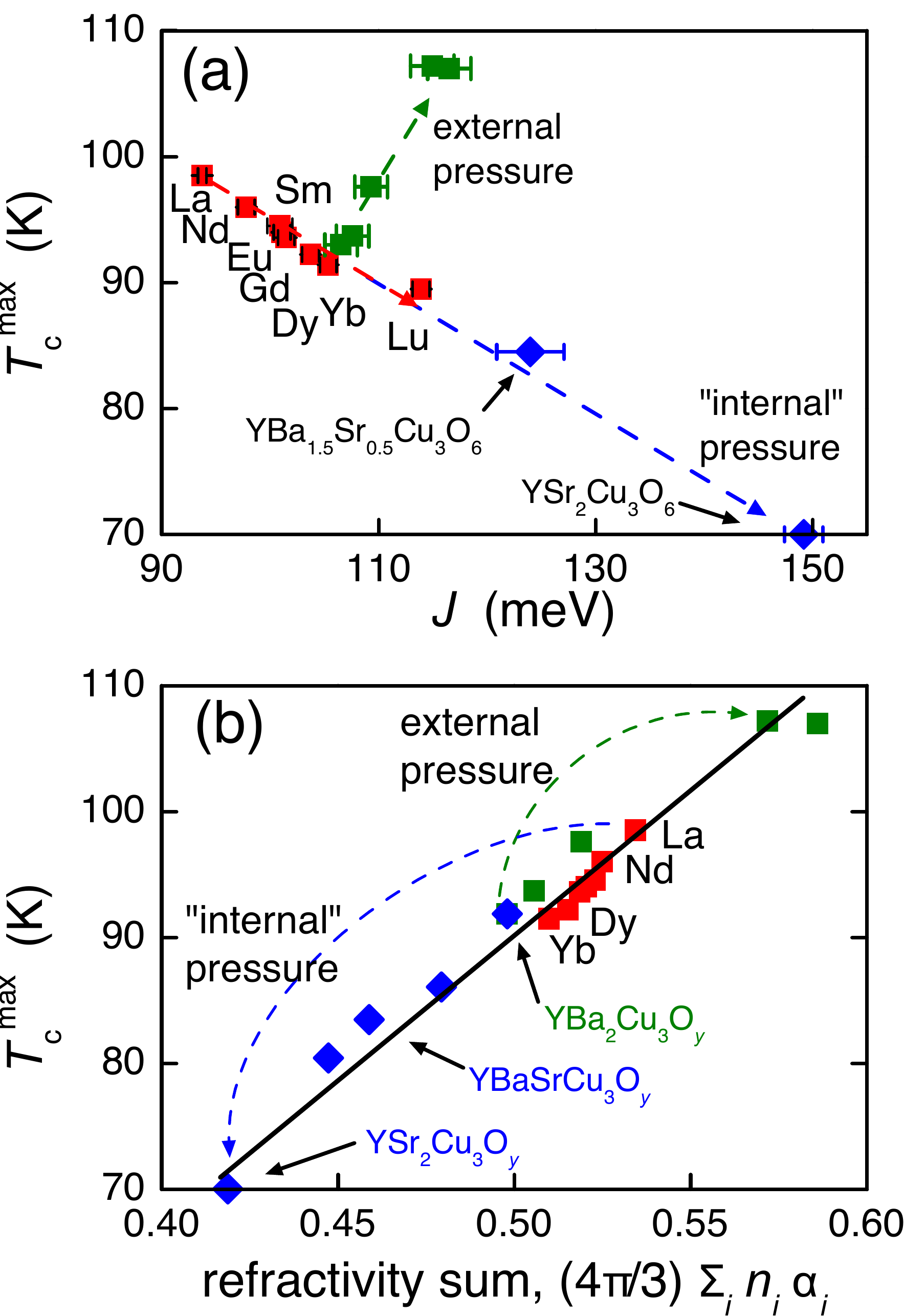}}
\caption{\label{fig:tcmax}(a) $T_c^{\textnormal{max}}$ plotted versus $J$ for single crystals of $R$Ba$_2$Cu$_3$O$_6$ (red squares) and for YBa$_{2-x}$Sr$_x$Cu$_3$O$_6$ (blue diamonds) with $x = 0.5$ and $2.0$. $T_c^{\textnormal{max}}$ anticorrelates with $J$ where ``internal pressure'' (blue trajectory) is the implicit variable. The green squares show that $T_c^{\textnormal{max}}$ versus $J$ under external pressure (green trajectory) is effectively orthogonal to the behavior for internal pressure.  
(b) $T_c^{\textnormal{max}}$ plotted against the polarizability sum $(4\pi/3)\sum_i{n_i\alpha_i}$ for $R$Ba$_2$Cu$_3$O$_y$ (red squares, $R$ = La, Nd, Sm, Eu, Gd, Dy, Yb) and YBa$_{2-x}$Sr$_x$Cu$_3$O$_y$ (blue diamonds, $x = 0, 0.5, 1.0, 1.25$, and 2) which summarize the effects of internal pressure. The green squares show $T_c^{\textnormal{max}}$ versus $(4\pi/3)\sum_i{n_i\alpha_i}$ for YBa$_2$Cu$_3$O$_y$ under external pressure, revealing a correlation which remains consistent with that for internal pressure.}
\end{figure}

We now plot in Fig.~\ref{fig:tcmax}~(a) $T_c^{\textnormal{max}}$ versus $J$ for the $R$-123 single-crystal series (red squares) using $T_c^{\textnormal{max}} = 98.5$ K for La-123 \cite{lindemer1994} and $T_c^{\textnormal{max}} =96$ K for Nd-123 \cite{veal1989} since these are the highest reported values of $T_c^{\textnormal{max}}$ in these cuprates (where $R$ occupation of the Ba site is minimized). Interestingly, $T_c^{\textnormal{max}}$ anticorrelates with $J$.

%This common behavior is perhaps surprising given that compression increases the nearest-neighbor hybridization, $t$. On the other hand reducing ion-size lowers the ionic polarizability, $\alpha_i$, which increases the on-site Hubbard $U$ through reduced screening \cite{brink1995}. In either case $J=-4t^2/U$ is increased but by apparently different mechanisms. However, compression also increases the density, $n_i$, of polarizable ions and it is the refractivity, $n_i\alpha_i$, that is the key variable as we shall see. This plays the same role for both internal and external pressure.

To move to yet higher $J$ values, we repeated the Raman measurements on a $c$-axis aligned thin film of YBa$_{1.5}$Sr$_{0.5}$Cu$_{3}$O$_6$ and on individual grains of polycrystalline YSr$_2$Cu$_3$O$_6$ prepared under high- pressure and temperature synthesis \cite{gilioli2000} [blue diamonds in Fig.~\ref{fig:tcmax}~(a)]. The anticorrelation between $T_c^{\textnormal{max}}$ and $J$ is preserved, but now out to a more than 60\% increase in the value of $J$. This is a very large increase and it is perhaps surprising that it is not reflected in the value of $T_c^{\textnormal{max}}$ if magnetic interactions alone set the energy scale for pairing.

To this plot we add data showing the effect of external pressure on $T_c^{\textnormal{max}}$ and $J$ in YBa$_2$Cu$_3$O$_7$ (green squares, for 0, 1.7, 4.5, 14.5 and 16.8 GPa).  The shift in $J$ with pressure is taken from Fig.~\ref{fig:jvsa} and the values of $T_c^{\textnormal{max}}$ at elevated pressures are from Refs \cite{mcelfresh1988, sadewasser2000}; see Ref.~\cite{mallettsm}. We note that the effects of external and internal pressure are orthogonal, highlighting the fact that the observed shifts in $T_c^{\textnormal{max}}$ simply do not correlate with $J$. This is our second main result.

These results differ from Ofer \textit{et al.} \cite{ofer2006} who reported correlations between $T_c^{\textnormal{max}}$ and the Ne\'{e}l temperature, $T_N$, in (La$_{1-x}$Ca$_x$)(Ba$_{1.75-x}$La$_{0.25+x}$)Cu$_3$O$_y$.  However, the effects they report are quite small compared with ours. Moreover, since $T_N$ is not directly related to $J$ their analysis required a model to estimate the exchange anisotropy and thereby convert $T_N$ to $J$.  We also note that this complex system has large nuclear quadrupole resonance linewidths reflecting a high degree of disorder. In our view the present study is more direct and reliable in its implications considering we have an ideal model system with constant buckling angle, an essentially pure dilatation strain tensor and relatively disorder free.  There may be some other systematic internal structural change which underlies our correlations but it is not yet evident.

\bog scattering only probes nearest-neighbor magnetic interactions \cite{singh1989} while recent RIXS studies \cite{guarise2010} reveal the presence of extended interactions involving next-nearest- and next-next-nearest-neighbor hopping integrals, $t'$  and $t''$ \cite{pavarini2001}. The additional extended exchange interaction is only about half the magnitude of the changes that we have imposed by ion-size variation. It is conceivable that inclusion of extended interactions might reverse the systematics reported here, however, in our view variations in $t'$ and $t''$  will have a stronger influence via the DOS by distorting the Fermi surface and shifting the van Hove singularity. RIXS studies of ion-size effects on $t'$ and $t''$ using our model system would settle this important question.

%Our observations suggest two possibilities, either that (i) the pairing energy scale is largely unrelated to $J$, or, (ii) pairing may be governed by magnetic interactions but other energy scales also vary across the R-123 series and have a much larger effect on $T_c^{\textnormal{max}}$ than $J$ does.%  These could include: $E_g$, $E_F -E_{\textnormal{vHs}}$, interplanar Josephson coupling, scattering rates, or condensation energy (governing fluctuations \cite{tallon2011}). Of these we consider it likely that increasing ion-size distorts the Fermi surface, shifting the vHs closer to $E_F$ by altering the relative magnitudes of the $t$, $t'$  and $t''$  hopping parameters \cite{pavarini2001}, thus raising the DOS. 

The contradictory behavior shown in Fig.~\ref{fig:tcmax}(a) contrasts the uniform simplicity of Fig.~\ref{fig:bvs} and suggests that some new element is needed to understand ion-size systematics. One arena where ion-size plays a key role is in the dielectric properties, where the ionic polarizability generally varies as the cube of the ion size \cite{shannon1993}.

The isolated CuO$_2$ planar array is electrostatically uncompensated and as such cannot constitute a thermodynamic system. It is necessary to include the compensating charges lying outside of the CuO$_2$ plane in any thermodynamic treatment and these will also mediate electron-electron interactions within the plane.  These charges reside on ions that are notably polarizable, resulting in the high dielectric constants observed in the cuprates \cite{reagor1989} and which will screen electronic interactions in the CuO$_2$ layers via incoherent fluctuations in polarization.  Additionally, there are \emph{coherent} excitations in such a medium giving rise to quantized bosonic polarization waves which can mediate $d$-wave pairing with a very large prefactor energy scale \cite{atwal2004, mallettsm}. %These charges reside on ions that are notably polarizable, resulting in the high dielectric constants observed in the cuprates \cite{reagor1989}. This can play a role through screening interactions via incoherent fluctuations in polarization. Additionally, there are coherent excitations in such a medium giving rise to quantized polarization waves.  Atwal and Ashcroft \cite{atwal2004} have shown that polarization-wave pairing in the $d$-wave channel in a highly polarizable medium can be dominant. 

In an early treatment of dielectric properties Goldhammer \cite{goldhammer1913} showed for sufficiently symmetric systems that the dielectric constant contains an enhancement factor $\left[ 1-\frac{4\pi}{3}\sum_i{n_i \alpha _i}\right] ^{-1}$ leading in principle to ``polarization catastrophe'' when $\frac{4\pi}{3}\sum_i{n_i \alpha_i} \rightarrow 1$. (The factor $\frac{4\pi}{3}$ is dependent on structure, the dielectric constant more generally being replaced by a dielectric matrix, and in a fuller treatment the enhancement factor is replaced by frequency- and momentum-dependent terms \cite{atwal2004}.) 

We focus first on the contributions from the noncuprate layers, and in Fig.~\ref{fig:tcmax}(b) we plot $T_c^{\textnormal{max}}$ vs $\frac{4\pi}{3}\sum_i{n_i \alpha _i}$ for $R$(Ba,Sr)$_2$Cu$_3$O$_y$ where the sum is over $R$, Ba, Sr, and the apical O(4) oxygens. Red squares summarize the effects of changing $R$ and blue diamonds the effects of replacing Ba by Sr. These are the internal pressure effects and the correlation is excellent. The polarizabilities are taken from Shannon \cite{shannon1993}.  If this correlation is to be meaningful it must resolve the paradox of the opposing effects of internal and external pressure. Qualitatively this seems possible because increasing ion size (decreasing internal pressure) increases the polarizability while increasing external pressure enhances the densities $n_i$, in both cases increasing the dielectric enhancement factor. %Moreover we can now resolve the paradox of the opposing effects of internal and external pressure. Increasing ion size (decreasing internal pressure) increases the polarizability while increasing external pressure enhances the densities $n_i$, in both cases increasing the dielectric enhancement factor. 
To test this we also plot $T_c^{\textnormal{max}}$ versus $\frac{4\pi}{3}\sum_i{n_i \alpha _i}$ for YBa$_2$Cu$_3$O$_7$ at 1 atm and 1.7, 4.5, 14.5 and 16.8 GPa (green squares) where we assume to first order that only the $n_i$, and not the $\alpha _i$, alter under pressure. For more details, see Ref.~\cite{mallettsm}. The correlation with polarizability is now preserved over a range of $T_c^{\textnormal{max}}$ from 70 to 107~K, including both internal and external pressure. This is our third main result and it partly explains the correlation with $V_+$ shown in Fig.~\ref{fig:bvs}.  The additional role of the apical oxygen bond length (which also contributes to the value of $V_+$) has yet to be clarified, but it may play a supplementary role in controlling the large polarizability of the Zhang-Rice singlet \cite{zhang1988} as distinct from controlling its stability as discussed by Ohta \etal \cite{ohta1991}.  

As shown in \cite{mallettsm} inclusion of the refractivities from the CuO$_2$ layers preserves the correlation with $T_c^{\textnormal{max}}$ but adds a further 0.4 to the refractivity sum bringing these systems close to polarization catastrophe. In practice this can implicate an insulator-to-metal transition or charge ordering, both of which are evident in the cuprates.

What may we conclude? These correlations might hint at a dielectric rather than magnetic pairing mechanism but it remains to be shown that the effects described here are not a proxy for some other structural systematics. And dielectric effects impact on magnetic interactions: a highly polarizable medium could merely be effective in screening long-range magnetic interactions \cite{raghu2012optimaltc}, or the on-site Hubbard $U$ \cite{brink1995}, by means of local \emph{incoherent} fluctuations. On the other hand, \emph{coherent} fluctuations of this medium, quantized waves of polarization, do mediate pairing on a large energy scale \cite{atwal2004} and might in fact be the elusive exchange boson. Both these distinct scenarios can be tested. Certainly the correlations reported here warrant deeper investigation.
%An increasing polarizability sum would then increase $T_c$, as observed.

Finally, based on our observed correlation and an inferred polarizability of $\alpha_{\textnormal{Ra}} = 8.3$ \AA$^3$ for the radium ion \cite{mallettsm} we predict $T_c^{\textnormal{max}} \approx 109 \pm 2$~K for YRa$_2$Cu$_3$O$_y$ and $117 \pm  2$~K for LaRa$_2$Cu$_3$O$_y$. A more amenable test is that $T_c^{\textnormal{max}}$ for Bi$_2$Sr$_{1.6}$La$_{0.4}$CuO$_6$ will be raised by substituting the more polarizable Ba for Sr, despite the introduction of additional disorder.

In summary, we describe a strategy that utilizes ion-size and external-pressure effects to measure how systematic changes in $U$, $t$, $t'$ and $t''$ relate to superconductivity in the cuprates. We apply this strategy to show that internal and external pressures have identical effects on the characteristic energy scale $J=-4t^2/U$ but opposite effects on $T_c^{\textnormal{max}}$, so that the latter does not correlate with the former. These results suggest that longer-range interactions play a significant role in defining the systematics of $T_c^{\textnormal{max}}$ in relation to structure and this role is governed by screening arising from the total system of core electrons, including the non-CuO$ _{2} $ layers.  This is supported by our finding that $T_c^{\textnormal{max}}$ correlates exceptionally well with the refractivity sum, for both internal and external pressures.  We suggest that a novel pairing mechanism involving coherent collective excitations associated with the ionic polarizabilities should be further explored.

%show that the characteristic spin-fluctuation exchange energy scale, $J$, increases as the ion size in $R$Ba$ _{2-x} $Sr$ _{x} $Cu$ _{3} $O$ _{y} $ decreases. We find this effective internal pressure has a quantitatively similar effect on $J$ to external pressure, but the opposite effect on $T_c^{\textnormal{max}}$. We find that $T_c^{\textnormal{max}}$ anti-correlates with $J$ when ion size is the implicit variable which suggests that longer range interactions play a significant role in defining the overall systematics of $T_c^{\textnormal{max}}$ as it relates to structure.  This role is governed by screening arising from the total system of core electrons, including the non-CuO$ _{2} $ layers.  This is reinforced by our finding that $T_c^{\textnormal{max}}$ correlates exceptionally well with the refractivity of the non-CuO$ _{2} $ layers, both for internal and external pressures.  But it could also signal an altogether different pairing mechanism involving coherent collective excitations associated with the ionic polarizabilities. 

This work is funded by the Marsden Fund (G.V.M.W., N.S., J.L.T.), the MacDiarmid Institute (B.P.P.M., J.L.T.) and the National Science Foundation, DMR-0907425 (N.W.A.).

\bibliography{C:/Users/MallettB/Dropbox/thesis/literature} %You need a file 'literature.bib' for this.



\section*{Supplementary material}

# Supplemental material for: Dielectric versus magnetic pairing mechanisms in high-temperature cuprate superconductors investigated using Raman scattering

B.P.P. Mallett[1,†], T. Wolf[2], E. Gilioli[3], F. Licci[3], G.V.M. Williams[1], A.B. Kaiser[1], N. Suresh[4], N.W. Ashcroft[5], J.L. Tallon[4]

[1]MacDiarmid Institute, SPCS, Victoria University P.O. Box 600, Wellington 6140, New Zealand
[2]Karlsruhe Institute of Technology, Inst. Solid State Physics, P.O. Box 3640, D-76021 Karlsruhe, Germany
[3]IMEM-CNR, Institute of Materials for Electronics and Magnetism, 43124 Parma, Italy
[4]MacDiarmid Institute, Industrial Research Limited, P.O. Box 31310, Lower Hutt, New Zealand
[5]Laboratory of Atomic and Solid State Physics, Cornell University, Ithaca, N.Y.14853-2501, U.S.A.

## 1. Sample preparation, processing and characterization.

High-quality $RBa_2Cu_3O_{7-\delta}$ single crystals were flux grown in Y-stabilized zirconia crucibles under reduced oxygen atmosphere, where necessary, to avoid substitution of R ions on the Ba site. The crystals were annealed for three days in flowing Ar at 600°C to remove oxygen from the chains then quenched to room temperature while still under Ar gas. We expect $\delta \geq 0.97$ from these annealing conditions and the measured mass change. Fully de-oxygenated chains are inert and this should eliminate any internal pressure-induced charge transfer that could otherwise occur between chain and $CuO_2$ layers. Moreover, even if there is some residual oxygen in the chains the doping state remains zero provided the oxygens remain isolated and chain segments do not start to form . Well-oriented films of $Y(Ba_{1-x}Sr_x)_2Cu_3O_{7-\delta}$ were synthesised by metallo-organic deposition on RaBITS[TM] textured Ni/W substrates with epitaxial buffer layers of $Y_2O_3/YSZ/CeO_2$. Polycrystalline samples of $YSr_2Cu_3O_{7-\delta}$ were prepared at 3 GPa and 1050°C using a multi-anvil apparatus with $KClO_3$ as an oxidant [2].

We use the room-temperature thermopower, $S(290)$, as a measure of doping state using the universal relationship reported by Obertelli, Cooper and Tallon [3]. To change doping state and $T_c$ we anneal samples in a vertical tube furnace under a flowing atmosphere of oxygen and nitrogen mixed in the desired proportions. Following equilibration the sample is quenched out, if necessary into liquid nitrogen, while still under this flowing atmosphere. $T_c$ is then measured using a Quantum Design PPMS SQUID magnetometer and plotted as a function of $S(290)$ in order to located the maximum, $T_c^{max}$.

Raman measurements were made at room temperature using a LabRam confocal microscope Raman spectrometer in back-scattering $B_{1g}$ geometry. For all samples except the polycrystalline $YSr_2Cu_3O_6$ the $B_{2g}$, $A_{1g}+B_{2g}$ and $A_{1g}+B_{1g}$ scattering geometries were used to check we were indeed measuring two-magnon scattering. The 514.5 nm line from an Ar ion laser with power $\leq 1$mW was focused to a spot of size ~1μm. A 300 lines/mm diffraction grating was used to capture the two-magnon peak in a single frame. For data in the inset to Fig. 2 we subtract a flat background, then normalise using the maximum in the scattering response.

In addition, samples of composition $YBaSrCu_3O_y$ were studied by muon spin relaxation at the Paul Scherrer Institute, Switzerland. We were concerned to check whether the disorder of Sr on the Ba site might play a role in weakening the superconductivity as a result of ion

substitution. But no significant suppression of superfluid density was observed despite the nominally high degree of disorder [4].

## 2. Strain tensor for internal and external pressure in $R(Ba_{1-x}Sr_x)_2Cu_3O_{7-\delta}$

The question arises as to how the strain tensor evolves in our model system with changing internal and external pressure. Are they equivalent? What of the associated shear strains? The strain tensor $\varepsilon_{ij}$ is often represented in matrix notation $e_i$, where the rules of transposition are outlined by J.F. Nye [5]. For orthorhombic and tetragonal systems the off-diagonal components $\varepsilon_{ij}$, where $i \neq j$, are zero and the key remaining strain components to consider are $e_1 = \varepsilon_{11} + \varepsilon_{22} + \varepsilon_{33}$ which is the dilatation, the tetragonal shear strain $e_2 = (2\varepsilon_{33} - \varepsilon_{22} - \varepsilon_{11})/\sqrt{3}$ and orthorhombic shear strain $e_3 = (\varepsilon_{11} - \varepsilon_{22})$. The other remaining components of the 1×6 strain matrix namely $e_4 = \varepsilon_{12}$, $e_5 = \varepsilon_{23}$, and $e_6 = \varepsilon_{31}$, are as noted zero. We have calculated these strain components for $R(Ba_{1-x}Sr_x)_2Cu_3O_{7-\delta}$ using $YBa_2Cu_3O_7$ at ambient pressure as the reference state using the structural refinements of Guillaume et al. [6], Licci et al. [7] and Gilioli et al. [2] to calculate the strain arising from internal pressure, and from Calamiotou et al. [8] under external pressure to 12.7 GPa. The strain $e_3$ is compared with $e_1$ in Fig. S1 and $e_2$ is compared with $e_1$ in Fig. S2

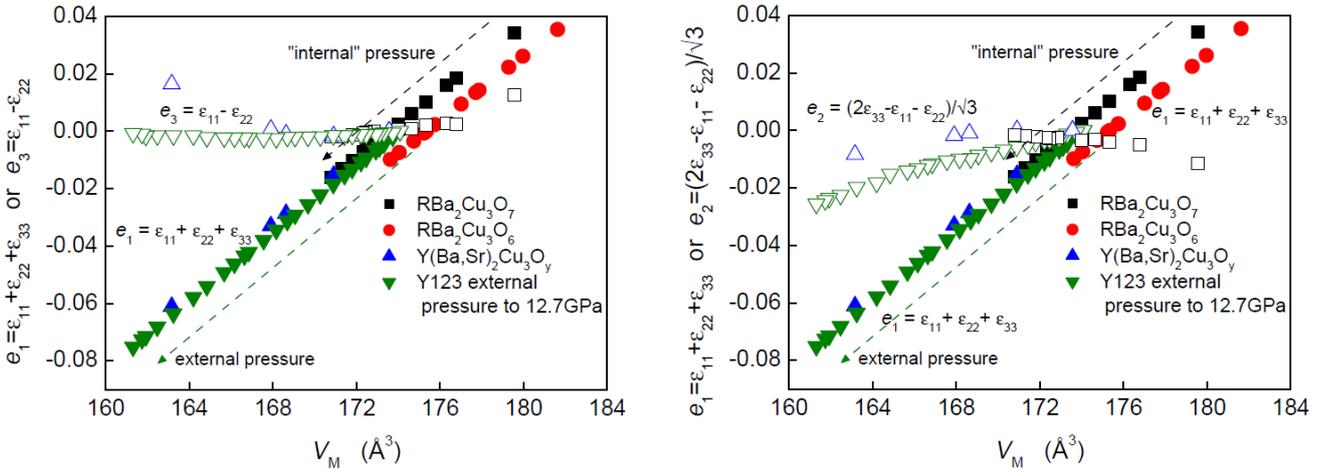

Figure S1 (LHS). The orthorhombic strain $e_3$ (open symbols) and dilatation strain $e_1$ (filled symbols) are plotted as functions of volume change arising from changing ion size (black, red and green symbols) and from increasing external pressure (green triangles).

Figure S2 (RHS). The tetragonal strain $e_2$ (open symbols) and dilatation strain $e_1$ (filled symbols) are plotted as functions of volume change arising from changing ion size (black, red and green symbols) and from increasing external pressure (green triangles).

The dilatation strain, $e_1$, follows the volume change in linear fashion across the entire range while the orthorhombic strain, $e_3$, remains close to zero across the range for both external and internal pressure, with the exception of the end points $LaBa_2Cu_3O_y$ (black open square) and $YSr_2Cu_3O_y$ (blue open up-triangle). These have diminished orthorhombicity arising from O5 occupancy in the former case and from incomplete oxygenation in the latter case. It is

presumed that fully oxygenated well-ordered chains will result in $e_3$ remaining close to zero in the ideal model system irrespective of the effective pressure is internal or external.

For the tetragonal deoxygenated system R(Ba$_{1-x}$Sr$_x$)$_2$Cu$_3$O$_6$ the shear strain $e_3=(\varepsilon_{11} - \varepsilon_{22})$ remains exactly zero under both internal and external pressure.

Turning to the tetragonal strain, $e_2$, shown in Fig. S2 it is evident that this also remains close to zero across the range as ion size is altered. The exceptions are the end members for the same reasons given above. This means that in the ideal model system the strain is essentially a pure dilatation as ion size is altered, whether at the rare-earth site or the alkali-earth site. (Close inspection shows that these have a small but opposite effect on $e_2$). To this we can add that across this model system there is negligible change in buckling angle [6,7].

Fig. S2 does however show a significant tetragonal strain arising from external pressure resulting from the higher compressibility along the *c*-axis. This only adds to the challenge of understanding the differences in effects between internal and external pressure because along with the decreasing tetragonality under external pressure (as shown in Fig. S2) the apical oxygen moves closer to the planar Cu site (see Jorgensen [9]) and, other things being equal, this is generally associated with a decrease in $T_c$. The opposite is observed as we note in our paper.

While it might loosely be argued that the somewhat different changes in tetragonal strain for external and internal pressure might possibly be associated with the different effects of these pressures on $T_c^{max}$, closer inspection shows that this cannot be the case. The last four members in Y(Ba$_{1-x}$Sr$_x$)$_2$Cu$_3$O$_y$ shown in Fig. S2 actually exhibit a *similar positive slope* in $e_2= (2\varepsilon_{33} - \varepsilon_{22} - \varepsilon_{11})/\sqrt{3}$ with $V_M$ as is found for YBa$_2$Cu$_3$O$_y$ under external pressure. In contrast, and as noted above, the internal pressure effect on $e_2= (2\varepsilon_{33} - \varepsilon_{22} - \varepsilon_{11})/\sqrt{3}$ for RBa$_2$Cu$_3$O$_y$ is small *but of opposite sign*. Yet in Fig. 4(a) of our MS the internal pressure effect on $T_c^{max}$ for these two systems is identical across the whole series while the external pressure effect is of opposite sign – one of our main results. Therefore, these tetragonal strain effects, while present to some degree, do not correlate at all with the external and internal pressure effects on $J$ and $T_c^{max}$.

### 3. Other studies

Our results might appear to be at odds with another Raman study where it is reported that, with increasing doping, the $B_{1g}$ gap parameter observed in low-frequency Raman scattering is proportional to $J$ [10]. But it is important to note that the $B_{1g}$ gap is not the SC order parameter, $\Delta_0$. For $B_{1g}$ probes anti-nodal regions of the Fermi-surface [11,12] and so will contain contributions from both the pseudogap, $E_g$, and $\Delta_0$, and in the under-doped region is completely dominated by $E_g$. Consequently, this reported correlation of the $B_{1g}$ gap with $J$ does not imply a link between $J$ and $\Delta_0$ but, rather, establishes a direct correlation between $E_g$ and $J$ as has already been inferred from specific heat [13] and inelastic neutron scattering [14]. It is the low-frequency $B_{2g}$ gap feature which probes $\Delta_0$ [11,12] and this does not correlate with $J$.

### 4. Choice of doping state

It has previously been found that $J$ is quite strongly doping dependent, in the case of Bi$_2$Sr$_2$Ca$_{1-x}$Y$_x$O$_{8-y}$ falling more or less linearly with doping from 125 meV in the undoped insulator to 45 meV in the lightly overdoped region at 0.20 holes/Cu [15]. Because of this strong doping dependence we focus on the undoped insulator (O$_6$) because this is the only

reproducible doping state for intercomparison of the different lanthanides. And perhaps more importantly, we wish to avoid pressure-induced charge transfer which can only be assured when the CuO chains are fully deoxygenated.

In the light of this, is the observed shift in *J* merely an inadvertent doping effect? We believe not, for the following reasons:

(i) As discussed above, for low oxygen concentration in the chain layer of R123, $\delta \approx 1$, the CuO2 planes are undoped and, in fact, insensitive to small changes in the oxygen concentration [16].

(ii) We have recently measured the pressure dependence of the thermopower of deoxygenated $Y_{1-x}Ca_xBa_2Cu_3O_6$ and there is no evidence of pressure-induced charge transfer [17]. Typically when $\delta < 1$ the application of pressure results in a strong decrease in the thermopower [18] reflecting an increase in doping state [3]. However when the samples are near fully deoxygenated ($\delta \approx 1$) the thermopower does become pressure independent.

(iii) If $\delta < 1$ then for a given $\delta$ the doping on the $CuO_2$ planes increases with decreasing ion size in the R123 system [19]. Given that *J* decreases as doping is increased [15], this would have the effect of *J* decreasing with decreasing ion size, the opposite to what we observe.

## 5. Relation between $T_c^{max}$ and *J*.

Many models for magnetic pairing, including the basic single-band Hubbard model, lead to the magnitude of $T_c^{max}$ being governed to leading order in *J*. For example, in weak coupling the condensation energy $U_0 \propto T_c^2/J$ [20] while the magnetic contribution to $U_0$ is given by *J* multiplied by a weighted integral of the structure factor [21]. Thus $T_c \propto J$. Moreover, Scalapino has shown that in the one-band Hubbard model where $J = -4t^2/U$ (with *t* being the nearest-neighbour hopping integral and *U* the on-site Coulomb repulsion energy) $T_c$ scales with *t* when $U/t$ is roughly constant [22]. Thus $T_c$ scales with *J*. It is also evident in the strong-coupling limit that when the full magnetic dispersion is employed in an Eliashberg scheme to calculate $T_c$ one still finds that $T_c$ scales to leading order with *J* [23].

## 6. Spin-charge stripes at large *J* ?

There is also this question: do the large values of *J* in the strongly-compressed structures exemplified by $YBa_{2-x}Sr_xCu_3O_x$ merely promote spin-charge stripes similar to what occurs in $La_{2-x}Sr_xCuO_4$, thereby reducing $T_c$ despite a magnetic pairing scenario which should otherwise raise $T_c$. We believe this is demonstrably *not* the case. In Fig. S3 we reproduce Fig. 4(a) but now include the data for $La_{2-x}Sr_xCuO_4$. It remains a stark outlier, with *J* mid-range but $T_c^{max}$ = 39 K. For $YSr_2Cu_3O_x$ the value of *J* well exceeds that for $La_{2-x}Sr_xCuO_4$ yet $T_c$ remains high. This shows again that *J* is not the dominant player. We believe that our model system remains well behaved across the full range of *J*. Moreover, $La_{2-x}Sr_xCuO_4$ fits the correlation in Fig. 1 (as shown by the left-most data point) as does $YSr_2Cu_3O_x$. This shows that there is some feature other than the magnitude of *J*, but reflected rather by the value of $V_+$, that determines their respective $T_c^{max}$ values. We have suggested that it is the ionic polarizability that at least partly affects the size of $T_c^{max}$. Indeed, the parameter $V_+$ is governed by the stretch of the in-plane Cu-O bond and the Cu-apical-oxygen bond length. Both are controlled by the ion sizes in the block layers which in turn govern the ionic polarizability.

## 7. Pressure dependence of $J$ and $T_c^{max}$

At one atmosphere pressure $T_c^{max}$ for YBa$_2$Cu$_3$O$_x$ is close to 93 K. To find $T_c^{max}$ at higher pressures it is important to note that the application of pressure has two effects: (i) to raise the magnitude of $T_c^{max}$ and (ii) to promote pressure-induced charge transfer which increases the hole doping state. Thus to determine $T_c^{max}$ at elevated pressures one must investigate underdoped YBa$_2$Cu$_3$O$_x$. The closer is the system to optimal doping (on the underdoped side) the lower is the pressure needed to attain optimal doping and $T_c^{max}$ rises little above its one atmosphere value. But with increasing underdoping, higher pressures are required to reach optimum doping and $T_c^{max}$ is raised further. This applies until the 60 K plateau is attained at a doping level of $p \approx 0.125$, when this pattern is broken and very much lower values of $T_c^{max}$ are then encountered [24]. No reports have yet been presented of these systematics in small increments of doping, and this is a gap that needs to be filled. However there are sufficient reports at several doping states to confirm the essential pattern. We then have the following data: ($P$ = 1 bar, $T_c^{max}$ = 93 K); ($P$ = 1.7 GPa, $T_c^{max}$ = 93.7 K [24]); ($P$ = 4.5 GPa, $T_c^{max}$ = 97.6 K [25]); ($P$ = 14.5 GPa, $T_c^{max}$ = 107.2 K [26]); and ($P$ = 16.8 GPa, $T_c^{max}$ = 107 K [24]).

For each of these pressures the value of $J$ is determined for YBa$_2$Cu$_3$O$_x$ from the inset to Fig.3, interpolating on a simple proportionality basis. We obtain: ($P$ = 1 bar, $J$ = 106.5 meV); ($P$ = 1.7 GPa, $J$ = 107.6 meV); ($P$ = 4.5 GPa, $J$ = 109.3 meV); ($P$ = 14.5 GPa, $J$ = 114.9 meV); and ($P$ = 16.8 GPa, $J$ = 116.5 meV). From these, the values of $T_c^{max}$ are plotted versus $J$ (with $P$ as the implicit variable) using the green squares in Fig. 4(a).

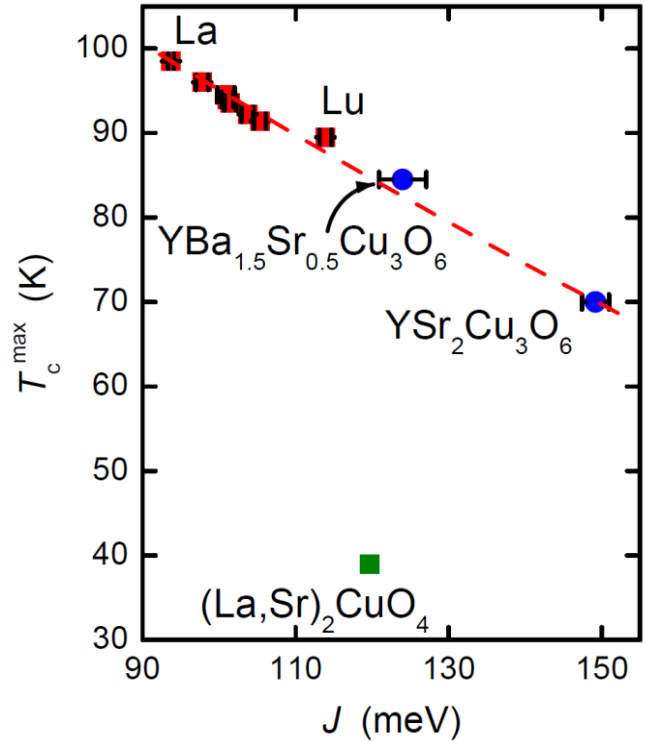

Figure S3. A reproduction (from Fig. 4(a)) of the ion-size effects on $T_c^{max}$ and $J$ in the model system RA$_2$Cu$_3$O$_x$ but now including La$_{2-x}$Sr$_x$CuO$_4$. The fact that the latter compound is a stark outlier when its $J$ value is only mid-range suggests that it is not the large value of $J$ which promotes "stripes" and suppresses $T_c^{max}$.

## 8. Pressure dependence of the refractivity sum

We assume to a first approximation that the effect of pressure is predominantly in the magnitude of $n_i$ which just scales as the density. In support of this we note that cations are found to display only a very weak dependence of their ionic polarizability on nearest-neighbor bond length [27]. Thus, defining $S = (4\pi/3) \Sigma_i n_i \alpha_i$, and $S_0 = S(P=0)$, then $S(P) = S_0 \times V_0/V = S_0 /(1 - P/B)$ where, again, $B$ is the isothermal bulk modulus. Because these calculations are applied at optimal doping we must use the bulk modulus for fully oxygenated $YBa_2Cu_3O_7$, namely, $B = 112$ GPa [9]. We obtain: ($P = 1$ bar, $S_0 = 0.4980$, $T_c^{max} = 93$ K); ($P = 1.7$ GPa, $S = 0.5056$, $T_c^{max} = 93.7$ K); ($P = 4.5$ GPa, $S = 0.5190$, $T_c^{max} = 97.6$ K); ($P = 14.5$ GPa, $S = 0.5718$, $T_c^{max} = 107.2$ K); and ($P = 16.8$ GPa, $S = 0.5860$, $T_c^{max} = 107$ K). These values of $T_c^{max}$ and $S(P)$ are plotted in Fig. 4(b) using the green squares which reveal a continuity with the data where internal pressure or ion size is the implicit variable.

## 9. Estimation of $T_c^{max}$ for A = Radium

The change in $T_c^{max}$ with ion size (either with Lu → La, or Sr → Ba) suggests that replacing $Ba^{2+}$ with $Ra^{2+}$ will continue this progression. The ionic polarizability of $Ra^{2+}$ does not seem to be reported in the literature so we simply used the fact that, to a good approximation, $\alpha_i \sim r_i^3$. In Fig. S4 we plot the polarizabilities for the alkali-earth ions as reported by Shannon [28] (the ionic radii are from Shannon [29]). The $\alpha_i \sim r_i^3$ dependence is apparent and suggests, if continued, that $\alpha_i(Ra^{2+}) = 8.03$ Å$^3$. Based on an extrapolation of Fig.4(b) this leads to a projected $T_c^{max}$ of $109 \pm 2$ K for $YRa_2Cu_3O_x$ and about $117 \pm 2$ K for $LaRa_2Cu_3O_x$.

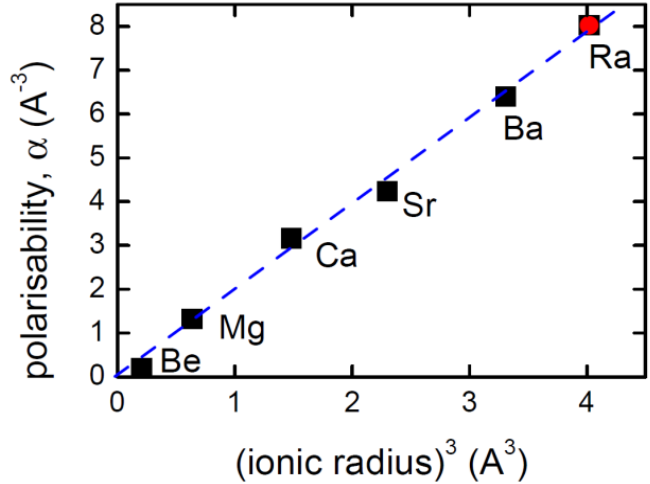

Figure S4. The ionic polarizability plotted versus the cube of the ionic radius in order to estimate $\alpha(Ra^{2+}) = 8.03$ Å$^3$. Polarizabilities are from ref. [28] and ionic radii from ref. [29].

## 10. Inclusion of the cuprate layers in the refractivity sum

The dielectric response of a cuprate must contain contributions from the frequency dependent polarizabilities of the total assembly of core electrons, including those of the copper-oxygen layers themselves. Initially, we adopted the approach of focussing on contributions from the non-cuprate layers since these dominate the dielectric properties; our primary interest was to isolate the effects of the added polarizable constituents. It is not entirely clear how any further

response attributable to the CuO$_2$ layers is to be incorporated in any simple fashion since (i) in the metallic state the polarization response of the copper and oxygen orbitals will be well screened by itinerant electrons, and (ii), metallicity arises by virtue of the fact that holes are doped into these very orbitals, and the corresponding hole concentrations are substantial. In our R(Ba$_{1-x}$Sr$_x$)$_2$Cu$_3$O$_7$ model system, for example, there are approximately 0.2 holes/Cu doped into each of the CuO$_2$ layers and 0.6 holes/Cu doped into the CuO chains (which are also conducting). Further, as indicated in our paper the doped holes reside on Zhang-Rice singlets which themselves, in non-itinerant states, have rather large polarizabilities.

For the sake of completeness (but cognizant of these complications), we have included in Fig. S5 all contributions to the refractivity sum by simply appealing to the polarizabilities for Cu$^{2+}$ and O$^{2-}$ reported by Shannon [28]. Evidently the relationship between $T_c^{max}$ and $(4\pi/3)\Sigma_i n_i \alpha_i$ is no longer linear as shown in Fig. 4(b) and it trends towards saturation. However $T_c^{max}$ still correlates with the refractivity sum, irrespective of whether the imposed pressure is external or internal. Notably, the refractivity sum is now approximately unity (up from ≈ 0.6 arising in the non Cu-O layers) and the system is interestingly close to the condition for a polarization catastrophe but, it must be emphasized, for an isotropic structure. A further interesting implication is that even in the undoped insulating state, where the CuO$_{1-\delta}$ chains are fully de-oxygenated and the refractivity sum then diminished by about 0.05, application of sufficient pressure ought to induce delocalization of some of the previously bound but polarizable charge (by approaching the condition $(4\pi/3)\Sigma_i n_i \alpha_i = 1$, but again, it must be stressed, for an isotropic system). Thus a pressure-induced insulator-to-metal transition in the undoped state becomes a possibility, and potentially more easily achievable in LaBa$_2$Cu$_3$O$_6$ than in YBa$_2$Cu$_3$O$_6$.

It appears, then, that approximate inclusion of Cu-O polarizabilities does not unduly influence the observed experimental trends and our major conclusions remain unaltered.

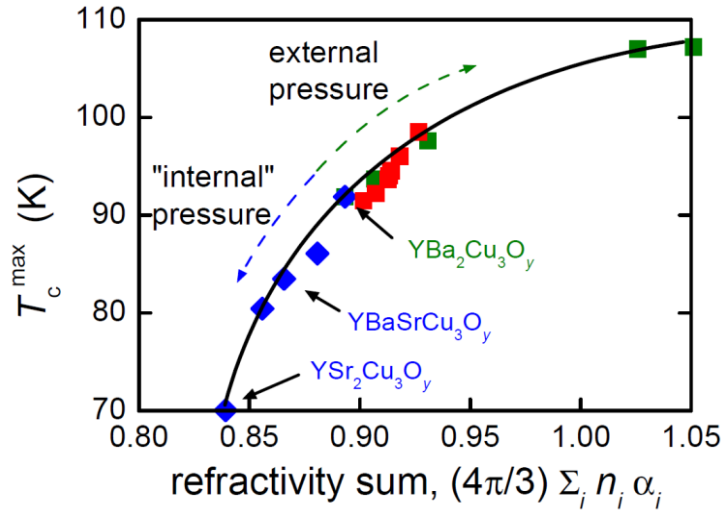

Figure S5. $T_c^{max}$ versus total refractivity sum for all ions including copper and oxygen from the CuO$_2$ and CuO layers.

## 11. Prefactor energy scale for polarization-wave induced pairing

The prefactor in any BCS estimate of $T_c$ where pairing is induced by exchange of bosonic polarization waves is the characteristic energy scale of the $E(\mathbf{k})$ dispersion for these polarization waves, an early example of which was given by Lundqvist and Sjölander [30]. Though the responding electrons are generally in localized states this is typified by a combination of physical constants which amount to an effective plasmon energy as made clear by the later polarization wave analysis of Lucas [31]. For the systems here this may be expected to be replaced by the equivalent quantity for the total assembly of core electrons. From these a comparison of the effective electron-electron coupling arising from polarization waves and phonons can be obtained (see equations (33) and (34) in Ref. [32]). It might be noted that the equivalent plasmon-like energy scale may be observable as an edge feature in the reflectivity occurring at 25-30 eV [33,34]; which then falls off at higher frequency as $\omega^{-4}$. It is more than two orders of magnitude higher than the nearest-neighbour exchange interaction and three orders of magnitude higher than the Debye energy relevant to pairing by exchange of virtual phonons.

## 12. References for Supporting Material:



\end{document}